\documentclass[amsmath,amssym,amsfonts,prl,manuscript]{revtex4}
\usepackage[dvips]{epsfig}

\begin{document}

\newcommand{\Xv}{{\mathbf x}}
\newcommand{\qv}{{\mathbf q}}
\newcommand{\av}{{\mathbf a}}
\newcommand{\cH}{{\cal H}}
\newcommand{\lB}{\ell_B}
\newcommand{\apa}{a_\parallel}
\newcommand{\ape}{a_\perp}
\newcommand{\Sv}{\hat{{\bf S}}}
\newcommand{\ev}{\hat{{\bf e}}}
\newcommand{\tv}{{\bf t}}
\newcommand{\nv}{\hat{{\bf n}}}
\newcommand{\bv}{{\bf b}}
\newcommand{\xv}{{\mathbf x}}
\newcommand{\uv}{{\mathbf u}}
\newcommand{\rv}{{\mathbf r}}
\newcommand{\grad}{\nabla}
\newcommand{\Wv}{{\mathbf W}}
\newcommand{\Q}{{\cal Q}}

\title{Chirality in Block Copolymer Melts:  Mesoscopic Helicity from Inter-Segment Twist }

\author{Wei Zhao}
\affiliation{Department of Polymer Science and Engineering, University of Massachusetts, Amherst, MA 01003, USA}

\author{Thomas P. Russell}
\affiliation{Department of Polymer Science and Engineering, University of Massachusetts, Amherst, MA 01003, USA}

\author{Gregory M. Grason}
\affiliation{Department of Polymer Science and Engineering, University of Massachusetts, Amherst, MA 01003, USA}

\begin{abstract}
We study the effects of chirality at the segment scale on the thermodynamics of block copolymer melts using self consistent field theory.  In linear diblock melts where segments of one block prefer a twisted, or cholesteric, texture, we show that melt assembly is critically sensitive to the ratio of random coil size to the preferred pitch of cholesteric twist.  For weakly-chiral melts (large pitch), mesophases remain achiral, while below a critical value of pitch, two mesocopically chiral phases are stable: an  undulated lamellar phase; and a phase of hexagonally-ordered helices.  We show that the non-linear sensitivity of meso-scale chiral order to preferred pitch derives specifically from the geometric and thermodynamic coupling of the helical mesodomain shape to the twisted packing of chiral segments within the core, giving rise to a second-order cylinder-to-helix transition.  
\end{abstract}
\date{\today}

\maketitle

Widely admired for their value to nanotechnology~\cite{hamley}, block copolymer (BCP) melts are also well-studied prototypes of soft-molecular assembly.  Even melts of the simplest, diblock architecture exhibit a rich spectrum of periodic, long-range ordered states, as a consequence of the competition between flexible chain entropy and demixing of unlike chemical domains~\cite{bates_fredrickson}. The nearly limitless flexibility of molecular design of copolymers, as well as a range of powerful experimental and theoretical methods for studying phase behavior, have provided an extensive platform for testing how changes in many factors such as molecular shape~\cite{grason_physrep, abetz} or the nature and range of inter-molecular forces~\cite{muthukumar_ober} dictate properties of long-range ordering in self-assembled systems.  One aspect of self-assembly that is crucial in a broader class of systems from liquid crystalline materials~\cite{harris, selinger_schnur} to biological assemblies~\cite{aggeli, grason_prl_07} is the influence of {\it molecular-scale chirality} on long-range ordering.  Despite the tremendous progress in understanding block copolymer thermodynamics in recent decades, the generic influence of chirality at the monomer scale on the meso-scale phase behavior is largely unknown.

Indeed, recent experimental studies by Ho and coworkers provide direct evidence that chirality at the scale of monomers has a profound impact on the meso-scale assembly of certain block copolymers~\cite{ho_04, ho_09, ho_12}.  Unlike the standard cylinder (C) phase of achiral copolymers, melts of polystyrene-{\it b}-poly(L or D)lactide (PS-PLLA or PS-PDLA) assemble into hexagonally-ordered arrays of helices, denoted as the H$^*$ phase, where the handedness of helices is shown to switch with a reversal of monomer chirality~\cite{ho_12}.  Even in this homochiral system, transfer of chirality from the molecular- to the meso-scale is not automatic:  at lower compositions of the chiral block, the achiral C phase is also observed~\cite{ho_09}.  These observations beg a number of unanswered questions about the influence of chirality on the block copolymer phase diagram.  Specifically, how does the mesoscopic packing of chains in the assembly couple to a thermodynamically preferred chiral packing in a way that reflects molecular handedness?  What key parameters dictate the stability of meso-chiral assemblies?  Finally, along with H$^*$, what other meso-chiral phases are stable in phase diagram of chiral block copolymer melts?

In this paper we investigate the equilibrium phase behavior of melt of diblock copolymers possessing a single chiral block.  Our approach is based on a recent self-consistent field theory (SCFT) approach to block copolymer melts that considers the mean-field cost for {\it gradients} in vector order parameter, ${\bf t}(\xv)$, the local mean segment orientation in ordered phases~\cite{zhao_12}.  The chiral nature of the polymers is modeled as a thermodynamic preference for a {\it cholesteric twist} of the orientational order associated with the chiral block.  We map the phase diagram of weakly-segregated melts in terms chiral segment composition $f$, degree of segregation and strength of chirality, as measured by $q_0 = 2 \pi/p$, where $p$  is the preferred pitch of the cholesteric helix of chiral segments.  Above a critical value of $q_0$, chirality dramatically alters the phase diagram leading to a stable H$^*$ phase, as well as new morphology, a phase of undulated layers, driven by cholesteric segment twist.  Finally, we analyze the C-to-H$^*$ transition based on both our SCFT results and a strong-segregation picture of chain packing to show that non-linear sensitivity of chirality transfer to $q_0$ derives specifically from the geometric interplay between mesodomain formation and inter-segment twist, giving rising to a {\it second-order phase transition} between these two phases.  

We consider incompressible melts of AB diblock copolymers possessing $N$ Kuhn segments, $f N$ of which belong to the chiral A block.  The total free energy of the melt (in units of $k_B T$) is written as,
\begin{equation}
F = \rho_0\chi  \int dV  \phi_{\rm A} (\xv)  \phi_{\rm B} (\xv) - S_{\rm chain} + F^* ,
\end{equation}
where incompressibility requires $\phi_{\rm A} (\xv) + \phi_{\rm B} (\xv)=1$.  The first term represents the mixing enthalpy of unlike segments, where $\phi_{\rm A} (\xv)$ and $\phi_{\rm B} (\xv)$ represent the local volume fractions of A and B monomers, $\chi$ is the Flory-Huggins parameter describing unlike segment interactions and $\rho_0^{-1}$ is the volume per segment.  The second contribution denotes the entropic free energy cost for polymeric chains to deviate from random-walk (Gaussian chain) statistics.   While these terms are familiar ingredients to the standard Gaussian-chain model of copolymer melts~\cite{matsen, fredrickson}, the third term represents the chiral coupling between gradients in the segment orientation and chain conformation.  Here, $\tv(\xv)$ is a vector order parameter corresponding to the local average of $\hat{\tv}_\alpha$, the orientation of the $\alpha$th A-block segment in the melt,
\begin{equation}
\tv(\xv) = \rho_0^{-1} \sum_{\alpha={\rm A}} \hat{\tv}_\alpha ~ \delta(\xv-\xv_\alpha) ,
\end{equation}
where the sum is carried out over all A segments on all chains~\cite{zhao_12}.  To describe the orientation-dependent interactions between the A segments in a coarse-grained way, we adopt a generalization of Frank elastic energy, standard to the theory of liquid crystalline order~\cite{degennes_prost}, describing square gradient costs for order-parameter variations in a cholesteric 
\begin{multline}
\label{eq: Ft}
F^*= \frac{\rho_0}{2} \int dV \Big\{ K_1 (\grad \cdot \tv)^2 + K_2 (\grad \times \tv)^2 + 2 q_0 K_2 \tv \cdot (\grad \times \tv) + K_2' \big[ \tv \cdot (\grad \times \tv)\big]^2 \Big\} ,
\end{multline}
where $K_1$, $K_2$ and $K_2'$ are elastic constants.  A lack of chiral symmetry at the segment scale corresponds to the case $q_0 \neq 0$, and a thermodynamic preference for non-zero cholesteric twist of segments, $\tv \cdot (\grad \times \tv)\neq0$~\cite{harris}.  In particular, for $K_2'=0$ the minimal free energy texture has pitch $p=2 \pi /q_0$.  While the first line of eq. (\ref{eq: Ft}) includes the square-gradient terms allowed by symmetry to second order in $\tv(\xv)$, it is necessary to include the higher order term ($K_2' \neq 0$) for stability.

\begin{figure}
\center \epsfig{file=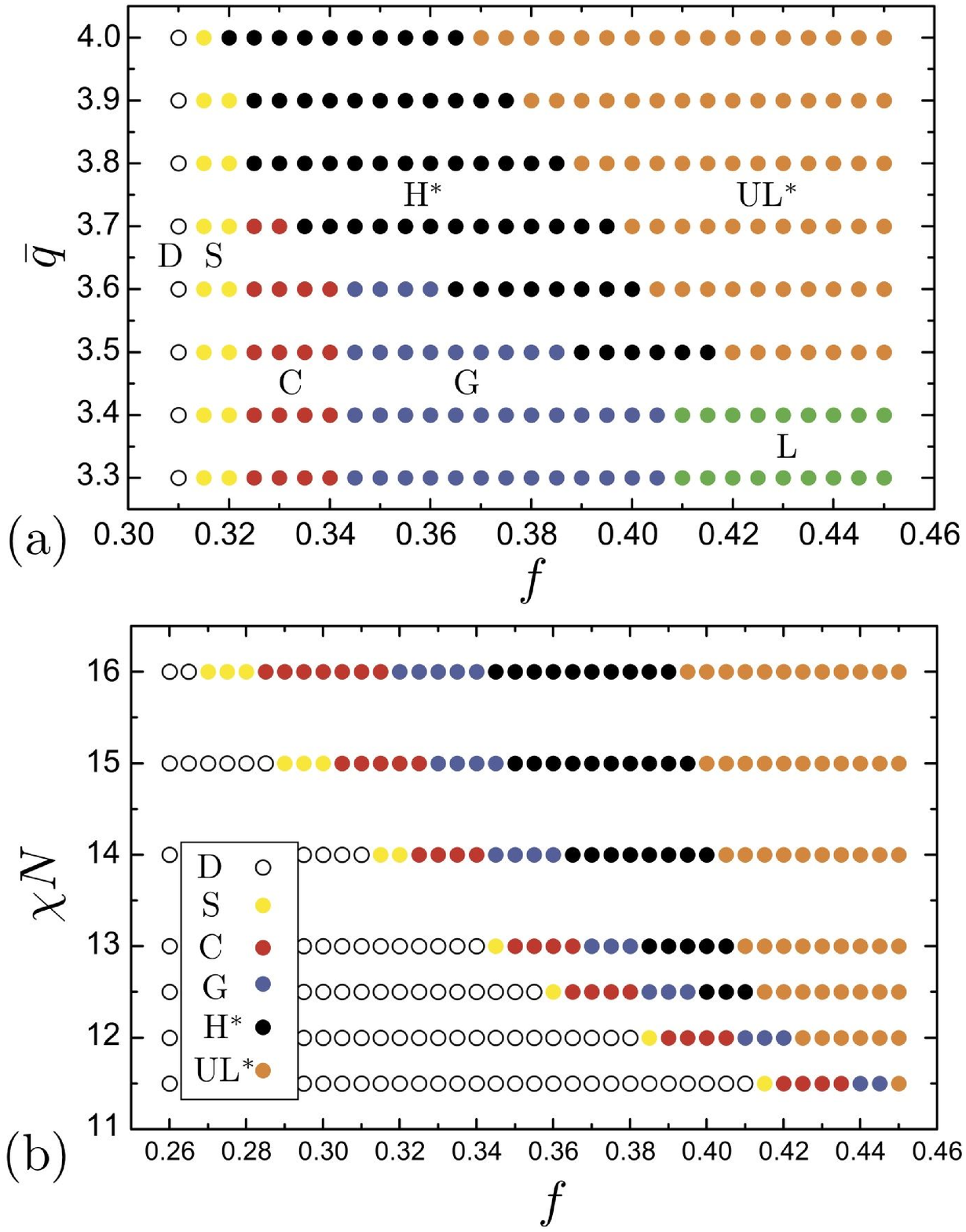, width=3in}
\caption{In (a), the phase diagram for fixed segregation strength, $\chi N = 14$.  In (b), the phase diagram a fixed degree of chirality, $\bar{q} = 3.6$.  Phases are labeled as in the text (D = disordered melt).  }
\label{fig: phase}
\end{figure}

The mean-field, SCFT analysis of copolymer melts possessing only density-dependent interactions is well-known~\cite{matsen}, and for brevity, we only summarize the key additional elements required by the self-consistent coupling to $\tv(\xv)$ (see ref. \cite{zhao_12} for details).  To determine composition and orientation profiles in the melt as well as the mean-field chain entropy in the ordered phases, segment-segment interactions are replaced by the {\it self-consistent} fields generated by their mean distributions.  We define $w_{\rm A}(\xv)$ and $w_{\rm B}(\xv)$ to denote the density-dependent fields acting on A and B blocks, respectively, and $\Wv(\xv)$ denotes the orientation-dependent field acting on the A block.  A modified diffusion equation describes the random-walk statistics on chain conformations subjected to the spatially varying fields,
\begin{equation}
\label{eq: q} 
\frac{\partial q}{ \partial s} = \left\{ \begin{array}{ll} \frac{N}{6} \big[ a \grad + \Wv(\xv)  \big] ^2 q - w_{\rm A} (\xv) q,  & s < f \\ 
\frac{N}{6} a^2 \grad^2 q - w_{\rm B} (\xv) q, & s > f \end{array} \right. ,
\end{equation}
where $q(\xv, s)$ describes the statistical weight of a section of chain spanning from the free end of the A block ($s=0$) to the $(s N)$th segment situated at $\xv$ and $a$ is the mean-segment size.  A similar equation describes the weight, $q^\dag(\xv, s)$, of ``diffusing" from the other free end of the chain ($s=1$) to the same point, such that the probability of the $(sN)$th segment of an {\it entire chain} at $\xv$ is $q(\xv, s) q^\dag(\xv, s) / \Q$, where $\Q = \int dV q(\xv, s) q^\dag(\xv, s)$ is the single-chain partition function.  The distributions $q(\xv, s)$ and $q^\dag(\xv, s)$ completely determine the mean composition profiles as well as the mean-orientational profile of the A-block segments according to
\begin{multline}
\tv (\xv) = \frac{V a}{6 \Q} \int_0^f ds ~ \Big[ q(\xv, s) \grad q^\dag (\xv, s)-  q^\dag(\xv, s)  \grad q (\xv, s) - 2a^{-1} \Wv(\xv) q(\xv, s) q^\dag(\xv, s) \Big] .
\end{multline}
The condition for mean-field in the melt requires that self-consistent fields are related to the mean composition and orientation profiles by $w_{\rm A} (\xv) - w_{\rm B} (\xv) = \chi [ \phi_{\rm B} (\xv) -  \phi_{\rm A} (\xv)]$ and $W_i(\xv)= \rho_0 \delta F^*/ \delta t_i(\xv)$.  These equations, along with the incompressibility condition, form a closed set whose solutions correspond to the equilibrium (mean-field) states of the melt.  The free energy of mean-field solution can be calculated directly from $\phi_{\rm A}$,  $\phi_{\rm B}$ and $\tv(\xv)$ and a relation for the entropy (per chain), $S_{\rm chain}/n =    \ln Q + V^{-1} \int dV [  w_{\rm A} \phi_{\rm A}+ w_{\rm B} \phi_{\rm B}+\Wv \cdot \tv]$. 

The self-consistent equations are solved numerically on a three-dimensional, periodic grid (of typical dimensions 20-35 on a side).  The distribution functions, $q(\xv,s)$ and $q^\dag(\xv,s)$, are solved by numerical integration of the ``diffusion + drift" equations, allowing for a direct calculation of $\phi_{\rm A}$,  $\phi_{\rm B}$ and ${\bf t}$ profiles for given self-consistent field distributions.  In turn, the self-consistency relations are solved for $w_{\rm A}$, $w_{\rm B}$ and $\Wv$ via an iterative, over-damped relaxation algorithm~\cite{fredrickson}.  To map the phase behavior, the minimal free-energy solution is identified for a given set of thermodynamic parameters. For this study, we consider melts described by the one Frank constant approximation, $K_1=K_2=N K_2'=K$, we set $N a^2 K =1/2$ and we focus on the regime where the chiral block is the minority component, $f < 0.5$.  Three thermodynamic parameters remain to govern melt behavior:  degree of segregation, $\chi N$; chiral fraction of the chain, $f$; and the {\it degree of chirality} $\bar{q} \equiv q_0 N^{1/2} a$, proportional to the ratio of (ideal) chain size to preferred cholesteric pitch of the chiral block.

\begin{figure}
\center \epsfig{file=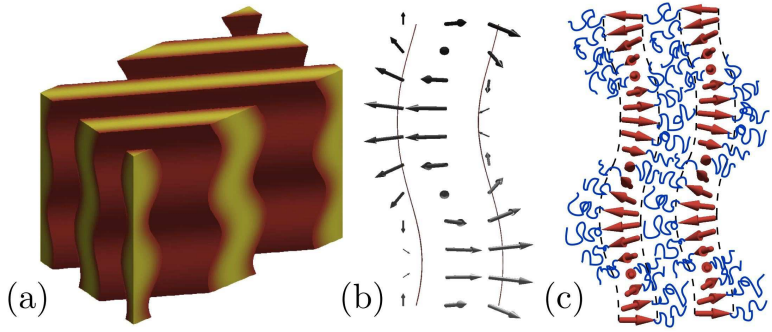, width=4in}
\caption{SFCT results for the UL$^*$ phase at $\chi N = 14$, $f=0.44$ and $\bar{q}=3.9$: in (a), the 3D volume of chiral domain is shown; in (b), an edge-on view of the segment orientation within the chiral layers (arrows indicate $\tv$-field orientation).  A cartoon schematic of the chain packing is shown in (c), with net polarization of chiral block segments shown as red arrows and achiral block shown as blue coil.}
\label{fig: UL}
\end{figure}

In Fig.~\ref{fig: phase}a we show the phase behavior in the $(f, \bar{q})$ plane for fixed, intermediate segregation strength $\chi N = 14$, while Fig ~\ref{fig: phase}b presents the $(f,\chi N)$ plane for a fixed degree of chirality, $\bar{q} = 3.6$.  In the former plane, we find that the phase diagram is divided into two regimes, weakly-chiral melts, $\bar{q}\lesssim 3.45$ and strongly-chiral melts, $\bar{q}\gtrsim 3.45$.  For weakly-chiral melts, we predict the standard sequence of BCP phases with increasing composition:  a BCC lattice of spherical domains (S), a hexagonal lattice of cylinders (C), a cubic double gyroid (G) and a lamellar phase (L).  Despite preference for twist in the system, the phases of the weakly-chiral regime are not chiral on the mesoscale, and consequently, the phase boundaries separating them do not show dependence on $\bar{q}$.  In contrast, in the strongly-chiral melts at the highest composition range we find that the achiral C, G, L phases are overtaken by two phases that do not appear in achiral BCP phase diagram, helical cylinders H$^*$, and an undulated lamellar phase UL$^*$.  From the $(f,\chi N)$ section of phase space at $\bar{q} =3.6$ in Fig.~\ref{fig: phase}b, we see that the stability of the mesochiral H$^*$ structure is not dictated by degree of chirality alone, as this morphology is found to have a window of thermodynamic stability only above a critical degree of segregation, $\chi N \simeq  12.5$.

\begin{figure}
\center \epsfig{file=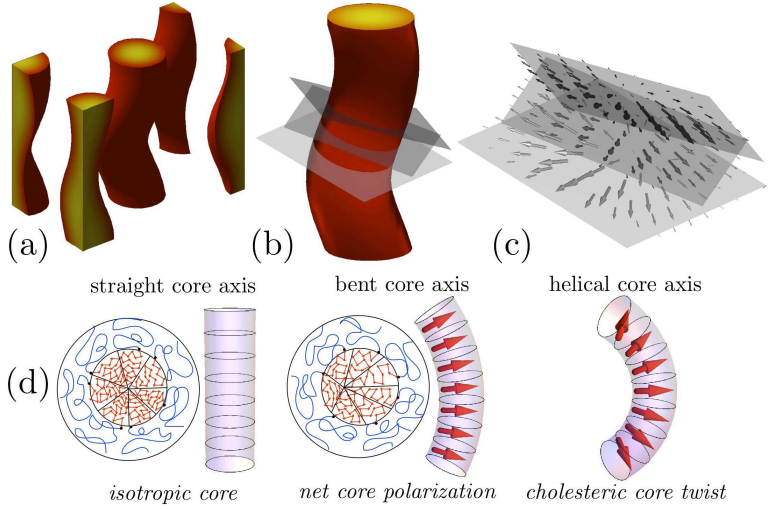, width=4in}
\caption{SFCT results for the H$^*$ phase at $\chi N = 14$, $f=0.345$ and $\bar{q}=3.6$: in (a), the 3D volume of chiral domain is shown; in (c) segment orientation within the consecutive cross-sectional slices through the chiral domain shown in (b).  In (d), a schematic depiction of coupling of segment orientation in cross section and the bending of twist of the backbone of the tubular core domain.  As in Fig. \ref{fig: UL}, the net polarization of the core block segments is depicted as a red arrow.  }
\label{fig: H}
\end{figure}

The thermodynamic stability of both the H$^*$ and UL$^*$ phases derives from the threading of cholesteric texture through the minority domains of either morphology, which is apparent upon inspection of the segment orientation distribution in these phases.  In the UL$^*$ phase Fig.~\ref{fig: UL}, $\tv(\xv)$ twists around a cholesteric pitch axis running {\it along the layer}, this direction also being the direction of lamellar layer undulation.  Unlike the standard ``double-layer" lamellar morphology, the UL$^*$ phase is composed of single layers whose chiral blocks are extended and ``highly-polarized".  The density of the achiral blocks emerging from either side of these layers oscillates due to the rotation of layer polarization, which in turn drives a periodic bending, or undulation, of lamella whose periodicity is locked to the cholesteric twist.  In H$^*$ (Fig.~\ref{fig: H}), we find the segments of minor (chiral) domain to be polarized along direction normal to the bending of the tubular domain, orienting towards the outside of the helix.  As this direction rotates along the pitch axis of the helix, overlaying consecutive sections of the helical core show a cholesteric twist that threads along the backbone of the helical domain (Fig.~\ref{fig: H}c).  

\begin{figure}
\center \epsfig{file=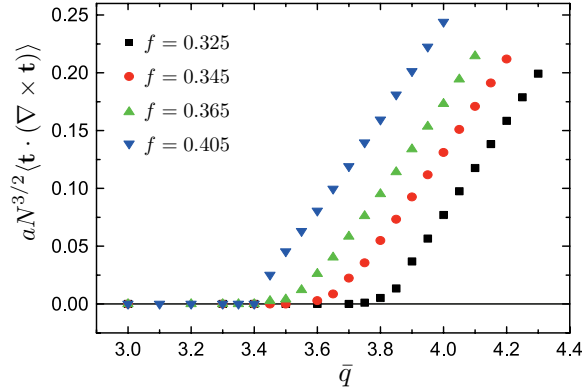, width=3in}
\caption{Volume-averaged cholesteric twist of SCFT results at $\chi N = 14$, showing a phase transition from C to H$^*$ with $\bar{q}$ and increase of $\bar{q}_c$ with decreasing chiral block composition.  }
\label{fig: twist}
\end{figure}

SCFT results demonstrate that chirality at the segment scale is a necessary, but not sufficient condition, for equilibrium stability of mesoschiral phases like H$^*$, consistent with experimental observations~\cite{ho_09}.  To explain the critical sensitivity of chirality transfer on $f$, $\chi N$ and $\bar{q}$ we consider a simplified analysis of the stability of the H$^*$ phase over C, the achiral cylinder phase.  At the heart of the chirality transfer from segments to mesoscale in the H$^*$ phase is the geometric coupling of {\it bending} of the tubular domain and {\it polarization} of segment orientation in the cross section, shown schematically in Fig.~\ref{fig: H}d.  Because the radial segment orientation within the core of the C mesodomain is isotropic---orienting equally in all directions in the planar section---a simple rotation, or twist, of consecutive sections generates no cholesteric twist, as measured by $\tv \cdot (\grad \times \tv)$.   Bending of the tubular domain breaks the isotropic symmetry of the C domain, tending to orient segments along the unit normal of the center line of the tube.  We adopt a strong-segregation perspective~\cite{milner} to estimate the degree of orientation by dividing the core domain into a series of wedges containing A-block chains extending radially from the centerline to the AB interface along the vector ${\bf R}_{\rm A}$, and note that the mean segment orientation within such as wedge is ${\bf R}_{\rm A}/(fN a)$.  Accounting for the excess volume of wedges on the outer side of the bent tubular domain, we average over all wedge orientations to find mean segment orientation $\tv\simeq -  \alpha (f) \kappa \nv R^2/(N a)$, where $\kappa$ and $\nv$ are the curvature and normal of the tube center line, $R$ is the outer domain radius of the tube, and $\alpha(f) = (3f^{-1/2}-1)/6 \approx f^{-1/2}/2$.  Hence, the helical rotation the normal direction along the backbone of the tubular domain, leads to a cholesteric twisting of the polarized segments in the core, $\tv \cdot (\grad \times \tv) \simeq - |\tv|^2 \tau$, where $\tau$ is the torsion of helix and we can estimate the chiral free energy gain (per chain) of the H$^*$ phase by replacing $\tv(\xv)$ with its cross-sectional mean,
\begin{equation}
F^*/n \approx - \frac{ K q_0^2 R^4 }{ N a^2} \kappa^2 ,
\end{equation}
where we set $\tau = q_0$ to maximize chiral free-energy gain.  Counteracting the tendency to polarize then twist the cores is the mechanical cost of bending the cylindrical microdomains (per unit cylinder length), $F_{\rm bend}/L = (B/2) \kappa^2$.  We estimate the modulus from the deformation cost of lamellar microdomains~\cite{ajdari} as $B = \beta(f)  \rho_0^{-1}N (\chi N)^{1/3} R^4$, where $\beta(f)$ is a numerical prefactor which we assume to be weakly dependent on $f$.  Combining these estimates we find the free-energy dependence of the $H^*$ phase in the small curvature limit,
\begin{equation}
F(H^*)/n \approx c (q_c^2 - q_0^2) \kappa^2 + O(\kappa^4) ,
\end{equation}
where $c \approx  \bar{K} R^4$ and $ \bar{K} \bar{q}^2_c \approx  (\chi N)^{-1/3}$ assuming the strong-stretching domain size $R \approx (\chi N)^{1/3} N^{1/2}a$.  Hence, the mechanical cost and chiral gain of helical buckling of core domain both grow with square curvature.  While for weakly-chiral melts, $q_0< q_c$, the straight ($\kappa =0 $) and achrial domains of the $C$ are stable, while for $q_0> q_c$ the core becomes thermodynamically unstable to helical bending.  Near to the critical degree of chirality we expect curvature of the core domain to grow {\it continuously} from zero as $\kappa \sim |q_0-q_c|^{1/2}$. Due to the geometric coupling between segment orientation and domain bending, we find that chirality transfer is highly non-linear process in tubular phases of copolymer melts, taking place as a second-order phase transition from the C-to-H$^*$ phase.  The signature of critical sensitivity to $q_0$ is shown in the volume average of cholesteric twist extracted from the SCFT, shown in Fig~\ref{fig: twist}.  Consistent with the thermodynamic argument above, we find no twist for $q_0<q_c$, while above the critical degree of chirality SCFT shows linear increase in cholesteric twist, following the critical scaling $\tv \cdot (\grad \times \tv) \sim \kappa^2 \sim |q_0 - q_c|$.

In summary, SCFT studies of chiral block copolymer melts reveal the equilibrium phase behavior of melts is critically sensitive to the thermodynamic preference for cholesteric twist of the chiral domain.  A drive for twist at the scale of chain segments indeed reshapes the structure and symmetry of the ordered phases on much larger length scales.  Importantly, the experimentally observed H$^*$ morphology is determined to be stable in the equilibrium phase diagram of chiral melts, and its thermodynamic stability derives directly from the geometric coupling of domain distortions and inter-segment twist.  While further studies are needed to map its full influence, there is little question that molecular chirality has a direct and potent impact on the structure and thermodynamics of block copolymer melts.

\begin{acknowledgments}
The authors are grateful to H. Johnston for extensive and invaluable discussions.  This work was supported by the MRSEC on Polymers at the UMass (WZ, GG) and DOE-DE-FG02-45612 (TPR).
\end{acknowledgments}

\end{document}